\documentclass[aps,prd,reprint,groupedaddress,onecolumn,10pt]{revtex4-1}
\pdfoutput=1 

\usepackage[T1]{fontenc} 
\usepackage{verbatim}
\usepackage{graphicx}
\usepackage{amsmath}

\def\be{\begin{equation}} 
\def\ee{\end{equation}} 
\def\bea{\begin{eqnarray}} 
\def\eea{\end{eqnarray}} 

\begin{document}

\title{The Wouthuysen Field Absorption Trough in Cosmic Strings Wakes}

\author{Oscar F. Hern\'andez}
\affiliation{Department of Physics, McGill University, \\
Montr\'eal, QC, H3A 2T8, Canada}
\affiliation{Marianopolis College, \\ 4873 Westmount Ave.,Westmount, QC H3Y 1X9, Canada}
\email{oscarh@physics.mcgill.ca}

\begin{abstract}The baryon density enhancement in cosmic string wakes leads to a stronger coupling of the spin temperature to the gas kinetic temperate inside these string wakes than in the intergalactic medium (IGM). The Wouthuysen Field (WF) effect has the potential to enhance this coupling 
to such an extent that it may result in the strongest and cleanest cosmic string signature in the currently planned radio telescope projects. Here we consider this enhancement under the assumption that X-ray heating is not significant. We show that the size of this effect in a cosmic string wake leads to a brightness temperature at least two times more negative than in the surrounding IGM. 
If the SCI-HI~\cite{Voytek:2013nua, Peterson:2014rga} or EDGES~\cite{Bowman:2012hf,EDGES2008}
experiment confirm a WF absorption trough in the cosmic gas, then cosmic string wakes should appear clearly in 21 cm redshift surveys of $z=10$ to 30. 
\\
\\
Accepted for publication in PRD.
\end{abstract}
\maketitle

\flushbottom

\section{Introduction}
\label{sec:intro}

Over the past years there has been a renewed interest in the possibility that
cosmic strings might contribute to the power spectrum of primordial
fluctuations. Many inflationary scenarios constructed in the context of supergravity models lead to the 
formation of gauge theory cosmic strings at the end of 
the inflationary phase \cite{Jeannerot1,Jeannerot2}, and in a large class of brane
inflation models, inflation ends with the formation of a network of cosmic superstrings~\cite{Sarangi} which can be stabilized as macroscopic objects in certain string models~\cite{Copeland}.
Finally, cosmic superstrings are also a possible remnant
of an early Hagedorn phase of string gas cosmology \cite{Brandenberger:2008nx}.
Whereas cosmic strings cannot be 
the dominant source of the primordial fluctuations~\cite{Albrecht,Turok},  
they can still provide a secondary source of fluctuations.
In all of the above mentioned scenarios, both a scale-invariant spectrum of
adiabatic coherent perturbations and a sub-dominant contribution
of cosmic strings is predicted.  In this sense, searching for signatures of cosmic strings is a way of probing particle physics beyond the Standard Model. By constraining the string tension $\mu$ we can constrain the particle physics symmetry-breaking pattern. 

The gravitational effects of the string can be parametrized by the dimensionless constant $G\mu$, where $G$ is Newtons gravitational constant. For cosmic strings formed in Grand Unified models, $10^{-8}<G\mu<10^{-6}$ whereas cosmic superstrings have $10^{-12}<G\mu<10^{-6}$~\cite{Brandenberger:2014lta}. Using combined data from the combined WMAP7 and SPT data sets, Dvorkin et al.~\cite{Dvorkin:2011aj} place an upper limit on the possible string contribution to the CMB anisotropy. In particular the power sourced by strings must be a fraction $f_{str} < 0.0175$ (95\% CL).
The Planck Collaboration~\cite{Ade:2013xla} has slightly improved this constraint to $f_{str}<0.01$ (95\% CL). Since $G\mu=1.3 \times 10^{-6}f_{str}^{1/2}$ this translates to a bound in terms of the string tension of $G \mu < 1.3 \times 10^{-7}$. Here and below, our limits on $G\mu$ are given at the 95\% confidence level.

It is interesting to characterize these upper limits in terms of the peculiar velocities generated by cosmic strings versus those generated by inflation.  The peculiar velocities induced by cosmic strings were studied by Brandenberger et al.~\cite{Brandenberger:1987un}. 
They found that in a model where all of the power
comes from strings (which requires $G \mu \simeq 10^{-6}$ to fit
the observed power spectrum), the rms velocities were of the
same order as in an inflationary model with the same total power.
This is easy to understand since the power spectrum of density
fluctuations from strings is scale-invariant like that produced
by inflation.
Since the velocities generated by strings are proportional to $G\mu$, we can scale the velocity perturbations they calculated by $f_{str}^{1/2}$ and compare to those from inflations (see figures 1 and 9 in~\cite{Brandenberger:1987un}). We thus have that the velocity perturbations from strings relative to those from inflation must be less than 0.05. These velocity perturbations are dominated by the effects of cosmic string loops versus wakes and the volume affected is approximately the volume inside the ensemble of loops~\cite{Vilenkinbook}. 

The string tension can also be constrained through the timing of pulsars~\cite{Vilenkinbook}. 
The decay of cosmic string loops emits gravitational waves, leading to a stochastic background dependent on $G\mu$. By using the limits imposed on the stochastic gravitational wave background from the European Pulsar Timing Array~\cite{vanHaasteren:2011ni}, Sanidas, Battye, and Stappers have placed a conservative limit of $G\mu < 5.3 \times 10^{-7}$~\cite{Sanidas:2012ee}. This constraint is weaker than that provided by the CMB anisotropy because of our lack of detailed knowledge of cosmic string networks. In particular the size of cosmic string loops $\alpha$, the spectrum of the radiation that they produce and the intercommutation probability $p$ all influence the contribution of loops to the  gravitational wave background. The size of cosmic string loops is characterized by the dimensionless loop production size $\alpha$, the fractional size of the loops relative the the horizon size at formation. Loops are considered large if $\alpha > \Gamma G\mu$, where $\Gamma$ is the ratio of the power radiated into gravitational waves by loops to $G\mu^2$. Numerical simulations suggest $\Gamma\sim50$~\cite{Casper:1995ub}. 
The intercommutation probability is unity for field theory strings, but can be as small as $10^{-3}$ for cosmic superstrings~\cite{Jackson:2004zg}. 
This conservative limit on string tension quoted above is for $\alpha=\Gamma G\mu$ and $p=1$. 
Interferometer experiments such as the LIGO-Virgo Collaboration can also search for the gravity wave background from loops. However these constraints remain weaker than those obtained from the CMB anisotropies and the pulsar timing arrays~\cite{Aasi:2013vna,Abbott:2009ws,Sanidas:2012ee}. 

Sanidas, Battye, and Stappers~\cite{Sanidas:2012ee} obtain more stringent constraints from pulsar timing arrays if $p<1$ (ref.~\cite{Sanidas:2012ee} fig.~14) or
when the size of loops is large (ref.~\cite{Sanidas:2012ee} fig.~13).
For $p=10^{-3}$ a conservative constraint on the string tension is $G \mu < 2.8\times 10^{-9}$. This occurs for loop size $\alpha=\Gamma G\mu$.
The simulation in~\cite{BlancoPillado:2011dq} suggests that cosmic string loops are large with $\alpha\approx 0.05$.
For $\alpha \approx 0.05$ and $p=1$ the limit obtained is $G \mu < 8.8\times 10^{-11}$~\cite{Sanidas:2012ee}. However there is a discrepancy between ref.~\cite{Sanidas:2012ee} and ref.~\cite{Blanco-Pillado:2013qja} in this last constraint, where for the same loop size the later work obtains
$G \mu < 2.8\times 10^{-9}$. In ref.~\cite{Blanco-Pillado:2013qja} the authors comment on this discrepancy and state that "a precise comparison is difficult, since both our loop sizes and velocities differ from models they considered." Despite these uncertainties, future pulsar timing experiments, for example in the the Large European Array for Pulsars (LEAP) project, have the potential to improve current constraints on the string tension by several orders of magnitude~\cite{Sanidas:2012ee,Ferdman:2010xq}. 
However, to date, the best firm constraints on the string tension come from the CMB power spectrum and give 
\be
G\mu \lesssim 10^{-7}.
\ee

In previous work~\cite{Brandenberger:2010hn,Hernandez:2011ym,Hernandez:2012qs}, we studied the signature and angular power spectrum of cosmic strings in 21cm radiation maps at redshifts $z$ between 20 and 30 corresponding to the dark ages, before star formation and non-linear clustering set in. The simpler physics that exists during this epoch means that an observed deviation from expected  21 cm brightness temperature would be a clean signature of new physics.  As described in~\cite{Brandenberger:2010hn}, the 21cm signature of a cosmic string wake has a distinctive shape in redshift space. However these previous papers ignored the effects of ultraviolet (UV) radiation. Here we consider UV radiation, in particular the Wouthuysen Field (WF) effect. Not only is this a first step towards studying the signatures of cosmic strings at lower redshifts, but the WF effect has the potential to greatly enhance the cosmic string signal.

Before the first luminous sources produced a large enough number of UV photons,
the 21 cm spin temperature $T_S$ of the cosmic gas was determined by a competition between Compton scattering and collisions. Compton scattering couples $T_S$ to the CMB radiation temperature $T_\gamma$, whereas collisions couple $T_S$ to the much cooler kinetic temperature $T_K$ of the cosmic gas. In higher density regions such as string wakes, collisions will lower the spin temperature and lead to an enhancement in the 21 cm brightness temperature.  This enhancement can be large enough to give a signal above noise for a string tension $G\mu\gtrsim3\times10^{-8}$~\cite{Hernandez:2012qs}. 

In the presence of UV radiation
hydrogen atoms can change hyperfine state through the absorption and re-emission of Lyman-$\alpha$ photons in what is known as the Wouthuysen-Field (WF) effect~\cite{wouth1952,field1958}.
Once enough UV photons are produced by the first galaxies, these transitions will again couple $T_S$ to $T_K$ leading to a more negative brightness temperature. 

Galaxies may also produce X-rays which heat the cosmic gas, and eventually reionization begins. Since the details of the sources driving these events is uncertain, it is not known when the WF effect will occur. If it occurs before the IGM has been sufficiently heated, this will enhance the absorption signal in the brightness temperature. But if insufficient UV photons are produced, the cosmic gas may reach the radiation temperature before the spin temperature couples to it. It is an open question as to whether this does or does not occur and global 21 cm experiments such as SCI-HI~\cite{Voytek:2013nua, Peterson:2014rga} and EDGES~\cite{Bowman:2012hf,EDGES2008} may soon give us an answer. Here will will assume X-ray heating is not significant since our concern is to compare the absorption signal, assuming it does exist, in the cosmic gas to that coming from a cosmic string wake. 

Many works \cite{Chen:2003gc,Barkana:2004vb,Hirata:2005mz,Pritchard:2005an,Furlanetto:2006tf,Chen:2006zr,Fialkov:2013uwm,Mirocha:2013gvq} have calculated the 21 cm brightness temperature in different scenarios for the redshift range $10<z<30$. Our purpose here is to show that the physics that leads to an absorption trough in the brightness temperature somewhere in this redshift range, will lead to an even larger effect in a cosmic string wake. 

We begin by reviewing the 21 cm brightness temperature both in the IGM and in cosmic string wakes in section~\ref{Tb} and then approximating the possible size of the WF absorption trough. In order to calculate and compare the size of the absorption trough in a cosmic string wake versus the surrounding cosmic gas we need to model the production of UV photons from the first luminous sources. We do this in section~\ref{UV} and use this to calculate the Lyman alpha coupling coefficient $x_\alpha$. This permits us to calculate the effect of these photons on the brightness temperature. 
In section~\ref{dTbmeas} we further discuss the measurement of a wake's brightness temperature. 
We present the results of our calculation in section~\ref{results}. In section~\ref{sigfore} we discuss the signal versus the foregrounds for a global 21 cm measurement, and we explain why we are optimistic that if a WF trough of at least 100 mK exists, it will be measured. Finally we discuss our conclusions in section~\ref{conclusion}.

\section{The 21 cm brightness temperature of the IGM and string wakes}
\label{Tb}
As explained in \cite{FOB}, the observation strategy for the 21 cm line is to measure the brightness temperature difference, $\delta T_b(\nu)$, a comparison of the temperature coming from the hydrogen cloud with the 
``clear view'' of the 21 cm radiation from the CMB. 
\be  \label{dTb1}
\delta T_b(\nu) \, = \,
\frac{
T_\gamma(\tau_\nu)-T_\gamma(0)}{1+z} 
\approx 
{(T_S-T_\gamma(0)) \over 1+z} ~\tau_\nu ~~.
\ee
$\tau_\nu$ is the optical depth and is given by:
\be  \label{tau}
 \tau_{\nu}(s)  =  \frac{3 h c^2 A_{10} x_{HI} }{32 \pi   \nu k_B } \, {{n_H \Delta s \, \phi(s,\nu)}
\over T_S}
 \approx 2.6\times 10^{-12}~ {\rm mK cm^2 s^{-1}}~{{x_{HI}\, n_H \Delta s \, \phi(s,\nu)}
\over T_S}
\ee
where $A_{10}=2.85\times10^{-15}\,$s$^{-1}$ is the spontaneous emission coefficient of the 
21 cm transition, $x_{HI}$ is the neutral fraction of hydrogen, $n_H$ is the hydrogen number density, 
$\Delta s$ is the thickness of our hydrogen cloud, $\phi(s,\nu)$ is the 21 cm line profile, and $T_S$ is the spin temperature.
Hence,
\be  \label{dTb2}
\delta T_b(z) \, \approx \, [2.6\times 10^{-12}~ {\rm mK \, cm^2 \, s^{-1}}]~{1\over 1+z}~
\left(1-{T_\gamma \over T_S}\right)
{x_{HI} \, n_H \Delta s \, \phi(s,\nu)} \, . 
\ee
Up to this point the hydrogen cloud could be anything, the cosmic gas or a cosmic string wake. It is the combination $x_{HI}\, n_H \Delta s \, \phi(s,\nu)$ and $T_S$ that differ for each. For the cosmic gas the brightness temperature difference is~\cite{FOB}
\be \label{dTbCG}
\delta T_b(z) \ = \, [9~{\rm mK}]
{
(1+\delta_b)
x_{HI}
(1+z)^{1/2}
\over
\left(1+{\partial{v_{pec}}/\partial{r}\over H(z)/(z+1)}\right)
}
\left(1-\frac{T_\gamma}{T_{S}}\right)
\left({\Omega_b\over0.05}\sqrt{0.3\over\Omega_m}{h\over0.7}\right)
~.
\ee
$\Omega_b, \Omega_m$ are the baryon and matter fractions today, $\delta_b$ is the baryon density fluctuation, 
$v_{pec}$ is the peculiar velocity, and $\partial{v_{pec}}/\partial{r}$ is the gradient of the peculiar velocity along the line of sight. 

For the brightness temperature difference of a cosmic string wake a very similar result holds~\cite{Brandenberger:2010hn,Hernandez:2011ym,Hernandez:2012qs}
\be \label{dTbwake}
\delta T_b^{wake}(z) \ = \, {[9~{\rm mK}]\over\sin^2\theta} 
\frac{n_{HI}^{wake}}{n_{HI}^{bg}} 
{
(1+\delta_b^{wake})
x_{HI}^{wake}
(1+z)^{1/2}
\over
\left(1+{\partial{v_{pec}}/\partial{r}\over H(z)/(z+1)}\right)
}
\left(1-\frac{T_\gamma}{T_{S}}\right)
\left({\Omega_b\over0.05}\sqrt{0.3\over\Omega_m}{h\over0.7}\right)
~,
\ee
The main distinguishing feature is the $\sin^{-2}(\theta)$ factor which comes from the line profile $\phi(s,\nu)$. $\theta$ is the angle of the 21~cm ray with respect 
to the vertical to the wake (see fig~\ref{fig:invsin2}).
The derivation of this factor is given in appendix A of~\cite{Hernandez:2012qs}, but it can be understood as follows. 
$\theta=0$ corresponds to a wake perpendicular to the line of sight. It is the gradient of the velocity along the line of sight that result in a line profile which is equal to the inverse of the frequency difference: $1/(\Delta \nu)$. Hubble expansion in the wake involves only the two long length directions, the width has decoupled from the Hubble flow and is growing by gravitational accretion. Because of this 21 cm radiation reaching the observer throughout the entire width of the wake have the same frequency, hence the singular nature of the line profile.
The factor however does not lead to a divergence in a physical measurement of the brightness temperature since it cancels out for small $\theta$ when the resolution of the measurement is taken into account as we will further discuss in section \ref{dTbmeas}.

Observing 21 cm radiation depends crucially on $T_S$. When $T_S$ is above $T_\gamma$ we have emission, when it is below $T_\gamma$ we have absorption.
Interaction with CMB photons, spontaneous emission, collisions with hydrogen, electrons, protons, and scattering from UV photons will drive $T_S$ to either $T_\gamma$ or $T_K$. Since the times scales for these processes is much smaller than the Hubble time, the spin temperature is determined by  equilibrium in terms of the collision and UV scattering coupling coefficients, $x_c$ and $x_\alpha$, as well as the kinetic and colour temperatures $T_K$, $T_C$:
\be \label{TsTkTc}
\left(1-\frac{T_\gamma}{T_{S}}\right)
=
{x_c \over 1+ x_c+x_{\alpha}} \left(1-\frac{T_\gamma}{T_{K}}\right)
+
{x_{\alpha} \over 1+ x_c+x_{\alpha}} \left(1-\frac{T_\gamma}{T_{C}}\right)
\ee

The optical depth for Lyman alpha photons is given by the Gunn-Peterson optical depth $\tau_{GP}\approx 2\times 10^4 x_{HI} (z+1)^{3/2}$.
Before reionization is significant ($x_{HI}$ not small), the large $\tau_{GP}$ value means that $T_C$ is driven to $T_K$ of the IGM. For for the rest of this work we work with $x_{HI}$ close to 1 and we take $T_C\approx T_K$. Thus: 
\be \label{TsTk}
\left(1-\frac{T_\gamma}{T_{S}}\right)
=
{x_c + x_{\alpha} \over 1+ x_c+x_{\alpha}} \left(1-\frac{T_\gamma}{T_{K}}\right)
\ee
The collision coefficients $x_c= {C_{10} T_\star \over A_{10} T_\gamma}$ for cosmic string wakes were discussed and calculated in \cite{Hernandez:2012qs,Brandenberger:2010hn,Hernandez:2011ym}. ($C_{10}$ is the de-excitation rate per atom for collisions) We discuss the Lyman coupling coefficient $x_\alpha$ in section \ref{UV}. 

We can approximate the size of the Wouthuysen Field effect in the cosmic gas under the assumption that X-ray heating is negligible. Before the kinetic temperature of the cosmic gas is significantly heated and reionized, we can approximate $T_K\approx 0.02\,{\rm K}\,(1+z)^2$, $x_{HI}\approx 1$. With $T_\gamma =2.725\, {\rm K} \,(1+z)$ we have:
\be \label{dTbCGapprox}
\delta T_b(z) \, \approx \, [9~{\rm mK}]
(1+z)^{1/2}
~{x_c+x_{\alpha} \over 1+ x_c+x_{\alpha}}
\left(1-{136\over 1+z}\right)
~,
\ee
In eq.~\ref{dTbCGapprox} and for the rest of this paper, we ignore the peculiar velocities, baryon density fluctuations, and take $\Omega_b=0.05, \Omega_m=0.3, h=0.7$. 

If $x_c+x_{\alpha} \gg 1$ then $T_S \approx T_K$. At redshift $z\sim 30$ collisions are rare in the IGM except for higher density regions such as minihaloes. In the mean density regions such a condition will not be reached until the Wouthuysen-Field effect is saturated, i.e. $x_\alpha \gg 1$.
\be \label{dTbCGapprox2}
\delta T_b(z) \, \approx \, [9~{\rm mK}]
{
(1+z)^{1/2}
}
\left(1-{136\over 1+z}\right)
~,
\ee
We see that if the WF effect is saturated before the cosmic gas is heated, the 21 cm line would show a strong absorption, with $\delta T_b < -170$ mK for $z<30$. Once heating begins the kinetic temperature approaches the radiation temperature, this strong absorption disappears. 

\section{UV photons and the Ly$\alpha$ coupling}
\label{UV}

To calculate the brightness temperature absorption trough due to the Wouthuysen Field effect we first need the Lyman coupling $x_\alpha$ and to do that we need a model for the production of UV photons. The Lyman coupling coefficient can be written as~\cite{Chen:2003gc,Barkana:2004vb,Hirata:2005mz,Pritchard:2005an} :
\be
x_\alpha = {P_{10}(z) T_\star\over A_{10}T_\gamma(z)}= 1.805\times10^{11}~{\rm cm}^2~ 
{S_\alpha J_\alpha(z) \over z+1}
\ee
where $T_\star=0.06817$~K is the equivalent temperature of the energy splitting between the two hyperfine states, $A_{10}=2.85\times10^{-15}$s$^{-1}$ is the spontaneous emission Einstein coefficient, and $T_\gamma(z)=2.725$~K~$(1+z)$, is the photon temperature. $P_{10}(z)$ is the de-excitation rate per atom from the triplet to singlet hyperfine state: $P_{10}(z)=0.020564$~cm$^2$s$^{-1}$~$S_\alpha J_\alpha(z)$. 

$S_\alpha$ is a correction factor of order one that accounts for spectral distortions~\cite{Hirata:2005mz}. We use the approximation given in eq. 43 of ref. \cite{FOB}
\be
S_\alpha=\exp\left[-0.803 \left(\frac{T_K}{\text{Kelvin}}\right)^{-2/3} \left(  {\frac{\tau_{\text{GP}}}{10^6}}\right)^{1/3}
\right]
\ee
where $T_K$ is the kinetic temperature of the cosmic gas and $\tau_{\rm GP}$ is the Gunn-Peterson optical depth. We are interested in evaluating this for redshift $z$ below $30$ and before reionization is significant and so we take $T_K\approx 0.02 {\rm K} (z+1)^2$ and $\tau_{\rm GP}\approx2\times 10^4 (1+z)^{3/2}$. With this, for redshift between 10 to 30,
we see that $S_\alpha$ is approximately between
0.65 and 0.85.
%

$J_\alpha(z)$ is the average Ly$\alpha$ flux in units of cm$^{-2}$ s$^{-1}$ Hz$^{-1}$ sr$^{-1}$. It is given by~\cite{Hirata:2005mz,Pritchard:2005an}
\be
J_\alpha(z)  =  \sum_{n=2}^{n_{\rm max}} J_\alpha^{(n)}(z) 
\ee
where $J_\alpha^{(n)}(z)$ is the background from photons that originally redshift into the Ly$n$ resonance, $\nu_n=(1-n^{-2})\nu_{LL}$, and cascade down to Ly$\alpha$. 
\be
J_\alpha^{(n)}(z) = \frac{(1+z)^2}{4\pi} f_{rec}(n)
\int_z^{z_n} {\rm d}z' \frac{c}{H(z')}\epsilon(\nu'_n,z')
\ee
$\nu'_n=\nu_n (1+z')/(1+z)$ is the frequency at redshift $z'$ that redshifts into that resonance at redshift $z$, 
and
$z_n$is the largest redshift from which a photon can redshift: 
$(1 + z_n)/(1+z)= (1 - (n+1)^{-2})/(1-n^{-2})$.
The recycle fraction $f_{\rm rec}(n)$ is the fraction of Ly$n$ photons that cascade through Ly$\alpha$:  $f_{\rm rec}(2)=1, f_{\rm rec}(3)=0,f_{\rm rec}(4)=0.2609$, and monotonically increase thereafter levelling off to 0.359 for large $n$~\cite{Hirata:2005mz,Pritchard:2005an}.
Following \cite{Pritchard:2005an, Barkana:2004vb} we truncate the infinite sum at $n_{\rm max}=23$ to exclude levels for which the horizon lies within the H II region of a typical galaxy.

The emissivity $\epsilon(\nu,z)$ gives the number of photons emitted at frequency $\nu$ and redshift $z$ per comoving volume, per proper time, per frequency. 
\be
\epsilon(\nu,z) = f_\star \, \bar{n}_b^0 \, \epsilon_b(\nu) \,
\frac{d }{d t}f_{coll}(M_{min}, z(t)).
\ee
where $f_\star$ is the efficiency that gas is converted to stars in haloes, $\bar{n}_b^0=\Omega_b\rho_{crit}^0/m_H$ is the mean baryon number density today, $\epsilon_b(\nu)$ is the number of photons produced at frequency $\nu$ per frequency per baryon in stars, and $f_{coll}(M_{min},z)$ is the fraction of mass collapsed in haloes with mass $M>M_{min}$. 

\begin{figure}[htbp]
\includegraphics[height=6cm]{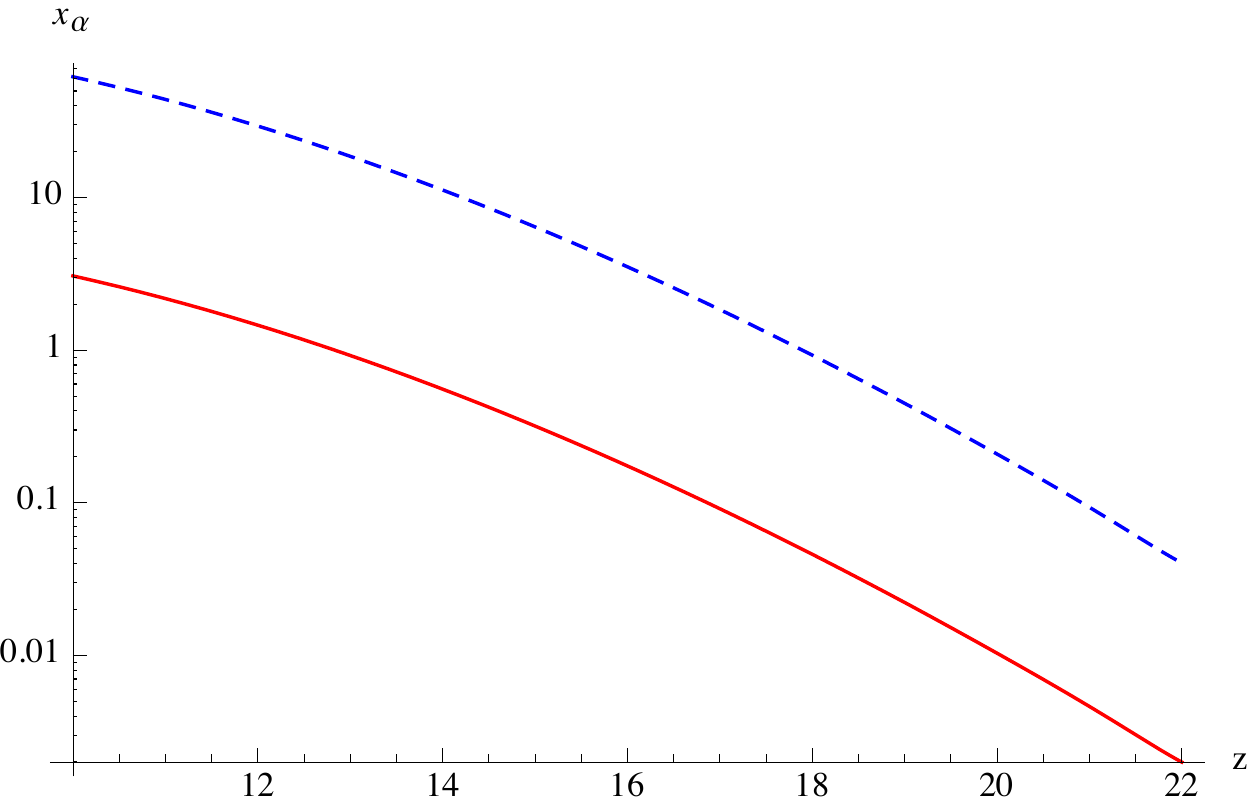}
\caption{The Lyman scattering coefficients $x_\alpha$ when UV photons are produced by Pop II (dotted blue) and Pop III (solid red) stars, where we take the star formation efficiency $f_\star=0.1$ and 0.01, respectively. }
\label{xalpha}
\end{figure}

The value of the efficiency $f_\star$ is a large source of uncertainty in our calculation and so our calculation of $x_\alpha$ will only be a rough guide to its value. The authors in~\cite{Barkana:2004vb,Hirata:2005mz,Pritchard:2005an,Furlanetto:2006tf,Chen:2006zr} use values of the efficiency between $10^{-3}$ to 0.1. We follow~\cite{Furlanetto:2006tf} and take $f_\star=0.1$ or 0.01 for Pop II or Pop III stars, respectively. Because the results presented in figure~\ref{xalpha} are proportional to $f_\star$, they can be rescaled if one uses other values of the efficiency.

In \cite{Barkana:2004vb} the emissivity $\epsilon_b(\nu)$ is taken as a separate power law in frequency between every pair of consecutive levels of atomic hydrogen so that the total Pop II stars emit 9690 and Pop III stars emit 6520 photons per baryon. We can approximate $\epsilon_b(\nu)$ as a constant equal to $9690/(\nu_{LL}-\nu_\alpha)$ or $4800/(\nu_{LL}-\nu_\alpha)$ for either Pop II or Pop III, and find better than 30\% or 6\% agreement, respectively, with the power law frequency dependence.

To determine $f_{coll}(z)$ we use the halo mass function 
$f_{ST}$ of Sheth \& Tormen~\cite{Sheth:1999mn} with the parameters given in~\cite{Chen:2006zr}:
\be
f_{coll}(m, z) = \int_{\delta_c(z)\over \sigma(m)}^\infty  d(\ln\nu) f_{ST}(\nu) 
\ee

We assume that the minimum mass $M_{min}$ is set by the virial temperature $T_{vir}\ge 10^4$~K, as in~\cite{Hirata:2005mz,Pritchard:2005an}, and we use the relationship between $M_{min}$ and $T_{vir}$ for a neutral gas given by:
\be
{M_{min}\over M_\odot}
= 1.05 \times 10^7 
\left[{T_{vir}\over10^4 {\rm K}} {21\over(1+z)} \right]^{3/2}
\left({0.3\over\Omega_m  }\right)^{1/2}
\left({0.7\over h} \right)
\ee
The time dependence in $f_{coll}$ occurs only through the redshift dependence of the linearized critical density $\delta_c(z)=\delta_c^0/D(z) \approx \delta_c^0 (1+z)$, where $\delta_c^0=1.686$ and $D(z)$ is the linear growth factor. Thus  
\be
\frac{d }{d t}f_{coll}(M_{min}, z(t))=(1+z) H(z) \frac{d }{d z}
f_{coll}(m, z) 
\biggr\rvert_{m = M_{min}} = H(z)f_{ST}({\delta_c(z)\over \sigma(M_{min})})
\ee
and
\be
J_\alpha^{(n)}(z) = {c\over4\pi} f_\star \, \bar{n}_b^0 \, \epsilon_b f_{rec}(n)
{\sigma(M_{min})\over\delta_c^0}
~(1+z)^2
\int_{\delta_c(z) / \sigma(M_{min})}^{\delta_c(z_n) / \sigma(M_{min})} {\rm d}\nu  \,   f_{ST}(\nu)
\ee

We now have everything we need to calculate the Ly$\alpha$ coupling $x_\alpha$. We do this for photons produced by Population II and Population III stars and present our result in figure~\ref{xalpha}.

\section{The wake's measured brightness temperature}
\label{dTbmeas}
It would appear from the $(\sin\theta)^{-2}$ factor in eq.~\ref{dTbwake} that there is a singularity at $\theta=0$ in the wakes brightness temperature. However if one considers the measured brightness temperature this is not so. 
\begin{figure}[htbp]
\includegraphics[height=8cm]{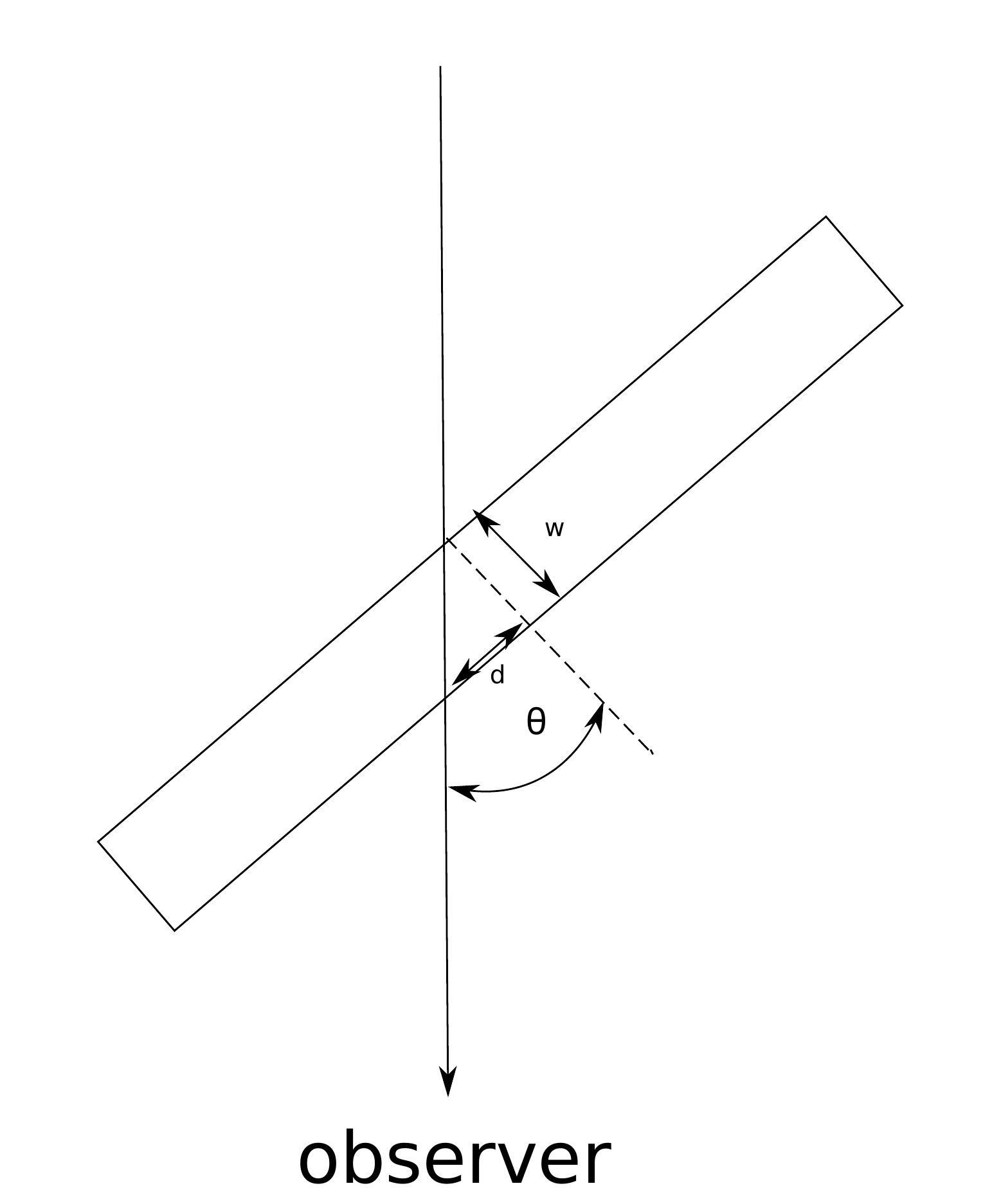}
\caption{A 21 cm light ray traverses a cosmic string wake of width $w$.} 
\label{fig:invsin2}
\end{figure}

As shown in fig.~\ref{fig:invsin2}, $\theta$ is the angle between the 21 cm ray reaching the observer and the normal to the wake. In a string wake only the planar directions expand in the Hubble flow, whereas the width grows by gravitational accretion, and hence any wake at a nonzero $\theta$ has a velocity gradient along the line of sight that depends on $\theta$. The relative velocity between the back and the front of the wake gives rise to a nonzero width of the 21 cm line and the line profile $\phi(\nu)$ is inversely proportional to this width. The brightness temperature, in turn, is proportional to the line profile. As $\theta$ goes to zero so does the line width, and hence the singularity in the line profile and brightness temperature. However any measurement of the 21 cm line involves a finite frequency resolution and so the measured brightness temperature shows no divergence. 

For small $\theta$, the frequency resolution of the measurement $\Delta \nu_{\rm res}$ will be greater than the frequency difference $\delta\nu_{\rm wake}$ from photons coming from the front and the back of the wake. Only at large angles will $\delta\nu_{\rm wake}$ be greater than $\Delta \nu_{\rm res}$.  

In an experiment the wake's measured brightness temperature is:
\be
[\delta T_b^{wake}(z)]_{\rm measured}=\int dz' ~W_z(z')~\delta T_b^{wake}(z') 
\ee
where $W_z(z')$ is a window function, peaked at $z$, that depends on the details of the experiment. We take $W_z(z')$ it to be a top hat function of width $\Delta z_{\rm res}$ centred at $z'$. 
The redshift resolution $\Delta z_{\rm res}$ is given by the frequency resolution of the measurement. For $\Delta z_{\rm res}$ is greater than the wake's redshift thickness $\Delta z_{\rm wake}$, we have,
\be
\label{dTbwakemeaz}
[\delta T_b^{wake}(z)]_{\rm measured}={\Delta z_{\rm wake} \over\Delta z_{\rm res}} \delta T_b^{wake}(z) + (1-{\Delta z_{\rm wake} \over\Delta z_{\rm res}} )\delta T_b^{IGM}(z)~~~~~~~\Delta z_{\rm res}>\Delta z_{\rm wake}
\ee
The redshift ratio 
${\Delta z_{\rm wake}/\Delta z_{\rm res}}$ is equivalent to the frequency ratio ${\delta \nu_{\rm wake}/\Delta \nu_{\rm res}}$ and hence we have
\be
\label{dTbwakemeanu}
[\delta T_b^{wake}(z)]_{\rm measured}={\delta\nu_{\rm wake} \over\Delta\nu_{\rm res}} \delta T_b^{wake}(z) + (1-{\delta\nu_{\rm wake} \over\Delta\nu_{\rm res}} )\delta T_b^{IGM}(z)~~~~~~~\Delta\nu_{\rm res}>\delta\nu_{\rm wake}
\ee
As shown in \cite{Hernandez:2012qs} 
\be
\label{dnu}
\delta\nu_{\rm wake}\, = \,{H(z) ~w~ \sin^2{\theta}  \over c~~ \cos\theta} ~~ \nu_{21}\, .
\ee
where $w$ is the wake's width. 
$\delta\nu_{\rm wake}$ increases monotonically in $\theta$ until $\theta$ reaches the value $\theta_1$ such that $\delta\nu_{\rm wake}(\theta_1)=\Delta \nu_{\rm res}$. Then for angles between $\theta_1$ and $\pi/2$, $[\delta T_b^{wake}(z)]_{\rm measured}=\delta T_b^{wake}(z)$. When this holds, we will get the strongest wake signal, since it will not be diluted by the cosmic gas as it is in eq.~\ref{dTbwakemeanu}

Let us see what frequency resolution we need to get a wide range of angles for which $[\delta T_b^{wake}(z)]_{\rm measured}=\delta T_b^{wake}(z)$.
We can use eq.~\ref{dnu} to find the value of $\sin^2(\theta_1)$ for which $\delta\nu_{\rm wake}(\theta_1)=\Delta \nu_{\rm res}$:
\be
\label{theta1}
\sin^2(\theta_1)=\mathcal{B}\sqrt{1+{\mathcal{B}^2\over4}} - {\mathcal{B}^2\over2}~~~~~~~~~~~~\mathcal{B}\equiv {\Delta \nu_{\rm res}\over\nu_{21}}{c\over w~H(z)}
\ee
The wake width $w$ is proportional to $G\mu(z+1)^{-1/2} H(z)^{-1}$ for shock heated wakes and to $G\mu(z+1)^{5/2} H(z)^{-1}$ for diffuse wakes~\cite{Hernandez:2012qs}.  For the small string tensions we are interested in ($G\mu\lesssim10^{-8}$) the wakes tend to be diffuse, and so
\be
\mathcal{B}={0.107 ~G\mu ~\over (v_s\gamma_s)^2}{\Delta \nu_{\rm res}\over 1~{\rm MHz}}{(z_i+1)^{1/2}\over (z+1)^{5/2}}
\ee
If we take $z=10,z_i=3000, (v_s\gamma_s)^2=1/3,G\mu=10^{-9},\Delta\nu_{\rm res}=0.01~{\rm MHz}$, we have that $\mathcal{B}=0.44$, and $\theta_1=0.36$ radians. For these parameters and the range of angles between 0.36 and $\pi/2$ radians we have that $[\delta T_b^{wake}(z)]_{\rm measured}=\delta T_b^{wake}(z)$. Decreasing $G\mu$ allows us to take a coarser resolution since the diffuse wake widens with decreasing string tension. At larger redshift $z$ the parameter $\mathcal{B}$ decreases and an even larger range of angles is possible. Thus we can evaluate our wake brightness temperature at a fiducial value of $\pi/4$ for comparison with the background IGM value.

\section{The brightness temperature evolution with Ly$\alpha$ photons}
\label{results}
With this in hand we calculate the brightness temperature for the cosmic gas and for cosmic string wakes (diffuse and shock heated). 
A cosmic string segment laid down at time $t_i$ (we are interested
in $t_i \geq t_{eq}$) will generate a wake with
physical dimensions:
\be  \label{size}
 l_1(t_i)\times l_2(t_i)\times w(t_i)~ =~ t_i ~c_1 ~\times~ t_i ~v_s\gamma_s ~\times~ t_i ~4\pi G\mu  v_s\gamma_s \, .
\ee
where $c_1$ is a constant of order one and $v_s\gamma_s$ is the speed time the Lorentz gamma factor of the string. After being laid down, the lengths Hubble expand whereas the wake width will grow by gravitational accretion. At a later time, parametrized by redshift $z$, a shock heated wake will have grown to physical dimensions. 
\bea
& & l_1(z)\times l_2(z)\times w(z)= \nonumber \\
& & \left( {3\over2}H(z)\sqrt{z_i+1\over z+1}\, \right)^{-3} 
       \left( c_1~\times~v_s\gamma_s \times 4\pi G\mu  v_s\gamma_s  {3\over10} {z_i+1\over z+1}\right)
\eea
where $z_i$ is the redshift that corresponds to time $t_i$. A diffuse wake will be wider by a factor discussed in eq.~3.2 of reference~\cite{Hernandez:2012qs}.
We take $c_1=1$ and $(v_s\gamma_s)^2=1/3$ and 
we restrict ourselves to the wakes laid down at at matter radiation equality, $z_i\sim3000$, since these will generically have the largest absorption brightness temperature~\cite{Hernandez:2012qs,Brandenberger:2010hn,Hernandez:2011ym}. 
We use eqs.~\ref{dTbwakemeanu},\ref{theta1} for the brightness temperature with the spin temperature given by eq.~\ref{TsTk}. 

\begin{figure}[htbp]
\includegraphics[height=6cm]{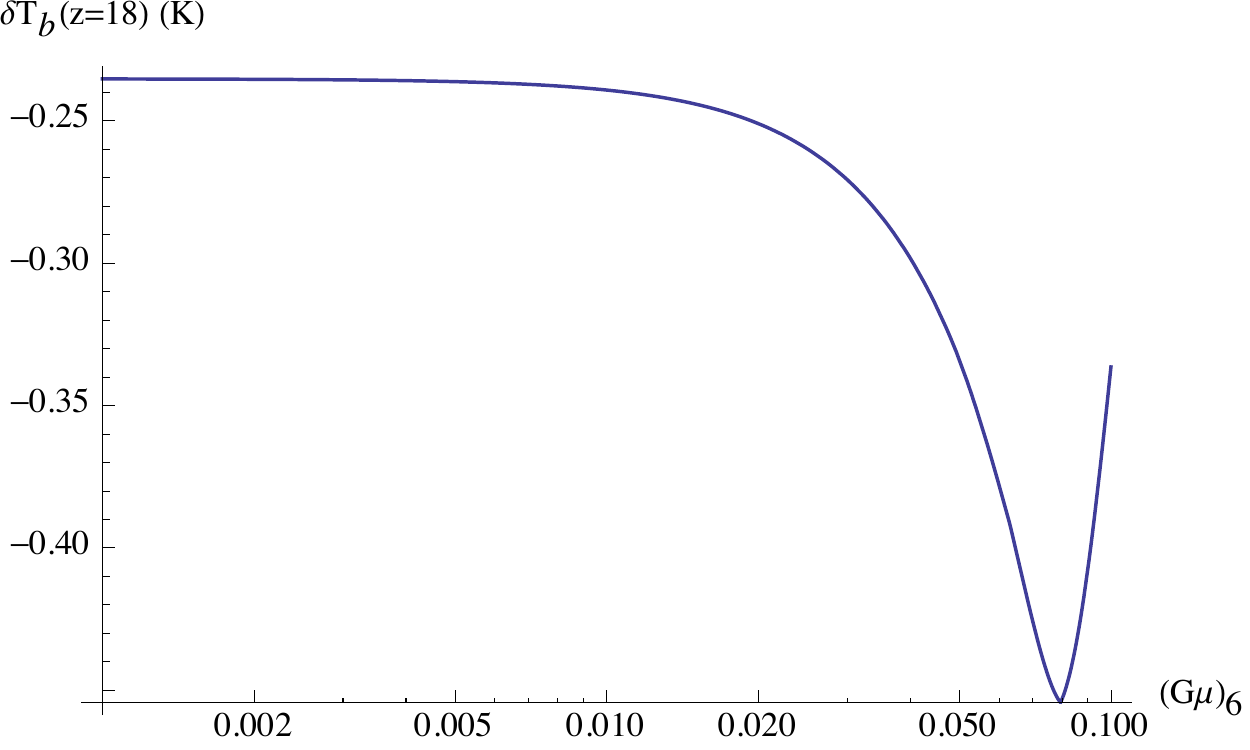}
\caption{The brightness temperatures (vertical axis) in degrees 
Kelvin with Pop II stars  at a redshift of $z=18$ as a function of the string tension $(G\mu)_6$ ($G\mu$ in units of $10^{-6}$). }
\label{GuII}
\end{figure}

\begin{figure}[htbp]
\includegraphics[height=6cm]{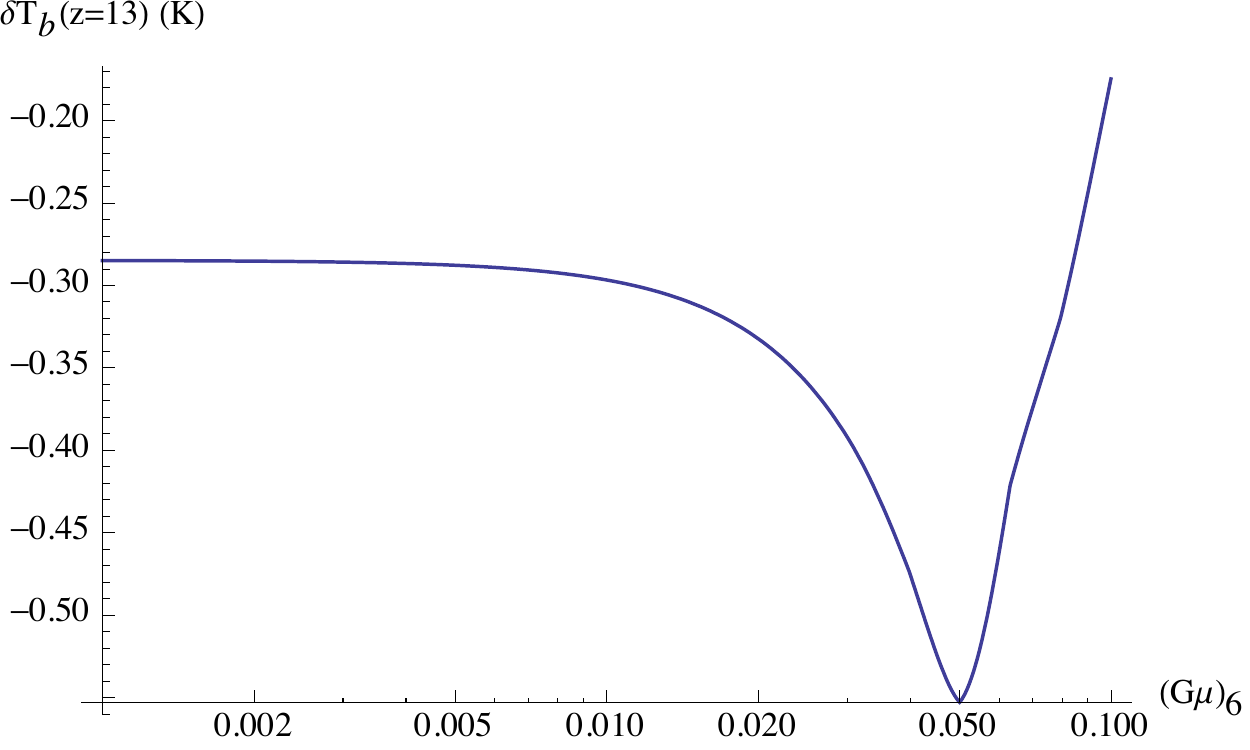}
\caption{The brightness temperatures (vertical axis) in degrees 
Kelvin with Pop III stars at a redshift of $z=13$ as a function of the string tension $(G\mu)_6$ ($G\mu$ in units of $10^{-6}$). }
\label{GuIII}
\end{figure}

The Wouthuysen Field effect couples $T_S$ to $T_K$ when $x_\alpha\approx1$. 
For Population II stars we find (see figure~\ref{xalpha}) that $x_\alpha\approx1$ at redshift $z\approx18$, and for Population III stars this occurs at about $z\approx13$. 
In figures~\ref{GuII} and \ref{GuIII} we plot the $G\mu$ dependence of the wake's brightness temperature for these cases.
Below $G\mu\lesssim10^{-8}$, the brightness temperature absorption trough plateaus at a value of approximately $-240$ mK with Pop II stars and $-290$ mK with Pop III stars whereas the brightness temperature of the IGM at $z=18$ and $z=13$ corresponding to the Pop II and Pop III stars is $-120$ mK and $-140$ mK, respectively. This plateau occurs because for a diffuse wake with small enough string tension, both the kinetic temperature and baryon density of the wake approach that of the cosmic gas \cite{Hernandez:2012qs}. The difference in brightness temperature is only due to the different line profiles arising from the different velocity gradients in a cosmic string wake versus the surrounding IGM. 

Figures~\ref{GuII} and \ref{GuIII} also show that the strongest signal occurs for $G\mu\approx 8\times 10^{-8}$ and $G\mu\approx 5\times 10^{-8}$ for Pop II and Pop III stars respectively. These values of $G\mu$ are largely determined by the shock heating condition $T^{wake}_K\gtrsim3\, T^{CG}_K$ which in turns gives a condition on the smallest $G\mu$ at a given redshift for which shock heating will occur:
\be
\label{Gmu1}
G\mu\gtrsim 1.6\times 10^{-9} (z+1)^{3/2} \, . 
\ee
When this condition is no longer met, our wakes becomes diffuse, with an increasing width but a decreasing overdensity. This occurs at $G\mu\approx 10^{-8}$ for redshifts between 13 and 18. As the string tension decreases even further the decrease in overdensity becomes more important than the increase in width and when $G\mu$ drops below $10^{-8}$ the brightness temperature plateaus, as we discussed in the previous paragraph.  

Finally, figures~\ref{GuII} and \ref{GuIII} show a decrease in the absolute value of the brightness temperature as the string tension continues to increase above $8\times 10^{-8}$ and $5\times 10^{-8}$ for Pop II and Pop III stars respectively. This is because the kinetic temperature in the shock heated wake increases as $(G\mu)^2$~(see \cite{Hernandez:2012qs,Brandenberger:2010hn}), and hence both the wake's kinetic and spin temperature approach the temperature of the CMB. 

\begin{figure}[htbp]
\includegraphics[height=6cm]{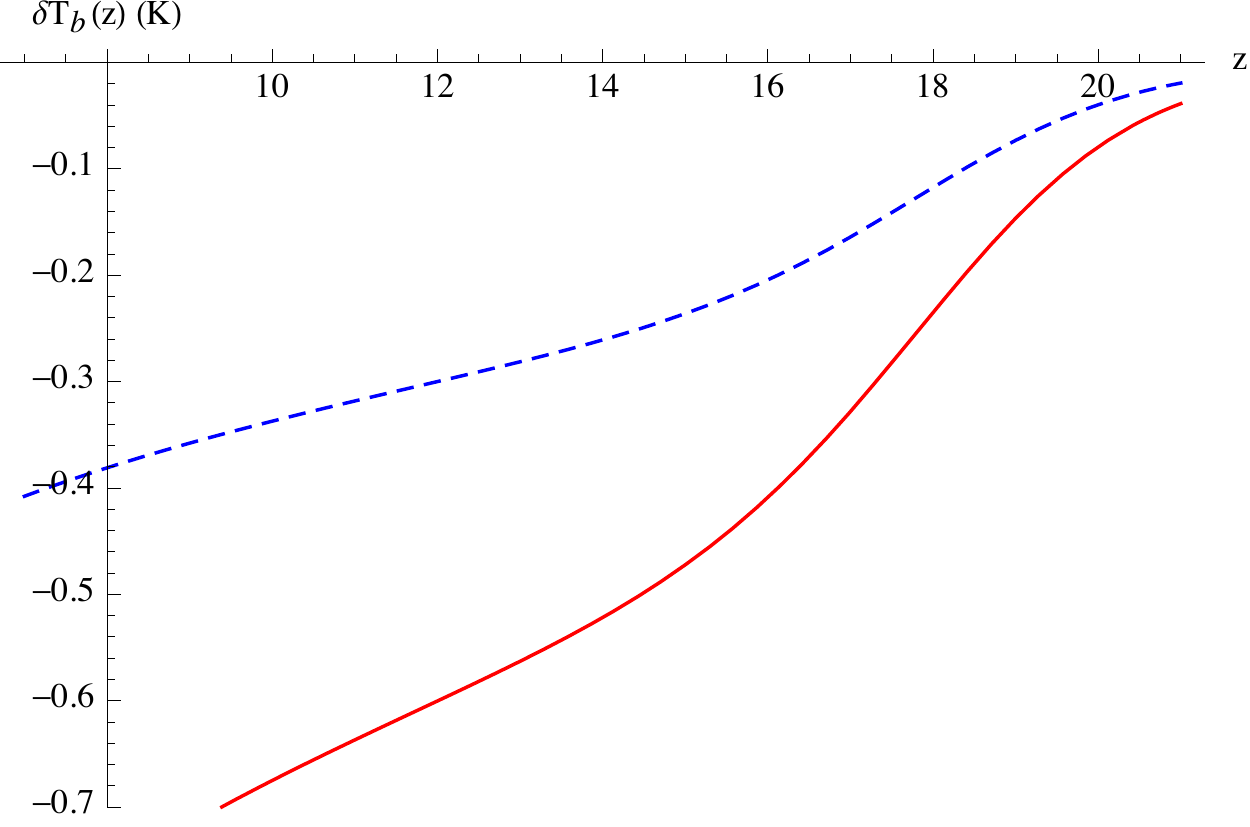}
\caption{The brightness temperatures (vertical axis) in degrees 
Kelvin as a function of redshift $z$ (horizontal axis) where the UV photons are produced by Population II stars. The surrounding cosmic gas is in dotted blue. A cosmic string wake with  $G\mu=10^{-9}$ is in solid red. }
\label{dTbII}
\end{figure}

\begin{figure}[htbp]
\includegraphics[height=6cm]{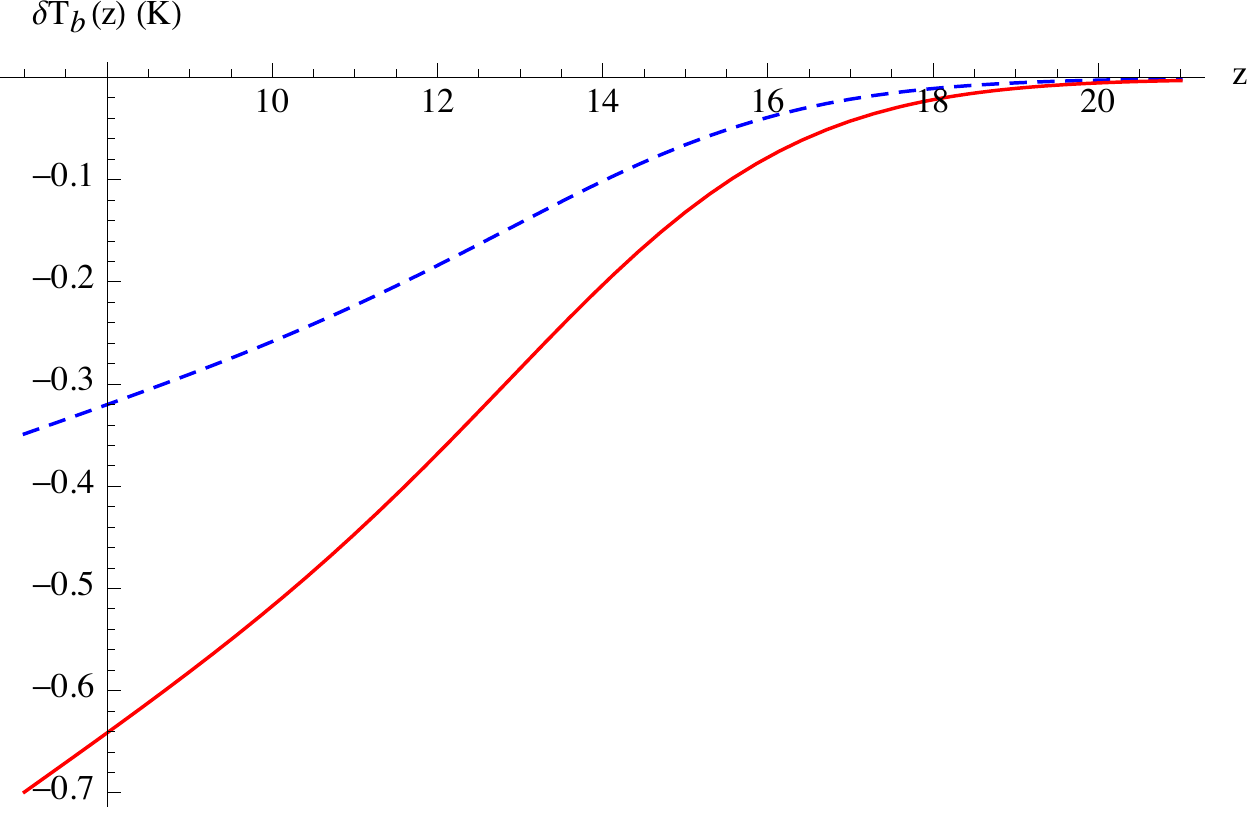}
\caption{The  brightness temperatures (vertical axis) in degrees 
Kelvin as a function of redshift $z$ (horizontal axis) where the UV photons are produced by Population III stars. The surrounding cosmic gas is in dotted blue. A cosmic string wake with  $G\mu=10^{-9}$ is in solid red. }
\label{dTbIII}
\end{figure}

In figures \ref{dTbII} and \ref{dTbIII} shows the amplitude of the expected temperature signal. There we plot the brightness temperature of the cosmic gas and of a cosmic string wake with string tension $G\mu\lesssim10^{-9}$ as a function of redshift. The absorption troughs rapidly become more significant at redshifts lower than those corresponding to an $x_\alpha=1$, i.e. $z=18$ or 13, for Pop II or Pop III starts respectively.  For example for Pop II stars at $z=16$, the IGM has a $\delta T_b(16)=-204$~mK, with the $\delta T_b^{wake}(16)=-410$~mK, a factor of two more negative. And for Pop III stars at $z=11$,  the corresponding numbers are -220 mK for the IGM, and -450 mK for the wake.  Even at redshifts where $x_\alpha<1$ there is a significant trough. For Pop II stars we have $\delta T_b(20)=-40$~mK, $\delta T_b^{wake}(20)=-80$~mK.  For Pop III stars we have $\delta T_b(16)=-40$~mK, $\delta T_b^{wake}(16)=-80$~mK.


\section{The Signal and the Foreground}
\label{sigfore}
The WF absorption trough we discussed here would occur at redshifts below $z=20$ and above $z=10$, i.e. frequencies between 70 to 140 MHz. As we scan this frequency range, the trough would be seen as a one hundred millikelvin step in the evolution of the global signal, which corresponds to the monopole of the brightness temperature~\cite{Shaver:1999gb}. Hence high angular resolution is not necessary and the global signal can be measured by a single dipole antenna. The problem with a global measurement at these frequencies are the foregrounds. Whereas the foregrounds for such a signal are very bright, they are expected to be smoothly varying in frequency. The rapid change in frequency for the cosmological signal versus the spectral smoothness of the foregrounds forms the basis for many of the foreground subtraction schemes that have been proposed. 

The authors of ref.~\cite{deOliveiraCosta:2008pb} have compiled a Global Sky Model of the radio sky from 10 MHz to 100 GHz using all available radio survey data. In ref.~\cite{Pritchard:2010pa} Pritchard and Loeb (PL) focus on the observations of a single dipole experiment antenna with 
a typical field of view of tens of degrees.
With such an antenna they found that they could fit the foreground temperature $T_{\rm sky}$, given by the Global Sky Model, to a polynomial in $\log(\nu)$ of not less than order 3. 
In particular for frequencies $\nu$ between 50 and 150 MHz they fit the sky temperature $T_{\rm sky}$ to:
$
\log T_{\rm fit}=\log T_0+ a_1\log(\nu/\nu_0)+a_2[\log(\nu/\nu_0)]^2+a_3[\log(\nu/\nu_0)]^3
$,
with $T_0=875 K$, $\nu_0=100$ MHz, $a_1=-2.47$, $a_2=-0.089$, $a_3=0.013$. 
The residuals visible after such a fit are dominated by numerical limitations of the Global Sky Model and had $\sqrt{\langle(T_{\rm sky}-T_{\rm fit})^2\rangle}\lesssim 1$~mK when averaged over the band. 

The analysis of PL now allows us to quantify how precisely we can measure the size of a 100 mK temperature dip in the WF trough. PL parametrize the 21 cm signal through 4 turning points which they name ${\bf x}_i=(\nu_i, \delta T_{bi})$ for $i=1,2,3,4$. Of particular interest for us here is their point ${\bf x}_2$ which gives the location and amplitude of the WF trough. They perform a Fisher matrix analysis on these four {\bf x} parameters which they then check with a Monte Carlo fitting for an experiment covering $\nu=40-140$ MHz in 50 bins, integrating for 500 hours and taking a third order polynomial fit for the foreground. The result of interest to us is given in their figure 12 where we can see that for a 1mK or 2 mK residual temperature, the 1 sigma on the measurement of the WF trough depth is 20 or 40 mK, respectively. In such a case a WF trough of order 100 mK can be both detected and distinguished from that due to a cosmic string wake at the several sigma level.  We should emphasize here that this analysis assumes that the instrument's frequency response can be calibrated out perfectly. Were this not the case, higher order polynomial would be necessary to fit the out the instrument's response. From PL's figure 12 we see that a 6th or 9th order polynomial fit to the foreground giving a 1 mK residual temperature would give a 1 sigma of 50 mK and 400 mK, respectively.  In this last case foreground fitting would be insufficient to measure the WF trough however we could then make use of other techniques as discussed in ref.~\cite{Liu:2012xy}. 

\section{Discussion and conclusion}
\label{conclusion}
We have seen that in the absence of significant heating from X-rays, the Wouthuysen Field effect leads to a large negative brightness temperature on the order of hundreds of millikelvin for the IGM and at least twice that for a cosmic string wake, even for a very small string tension. 
For small string tensions the wake temperature and the wake baryon density are not significantly different from that of the IGM, however they have decoupled from the Hubble flow and because of that the line profile of the 21 cm ray reaching the observer from a wake leads to a brighter brightness temperature. The enhancement in the brightness temperature relative to the cosmic gas is expressed through the $(\sin\theta)^{-2}$ factor in eq.~\ref{dTbwake}.

The WF absorption trough is even greater in shocked cosmic string wakes. There the higher density regions make collisions more important than in the cosmic gas, and they are also hotter. However shocked wakes tend to occur for string tensions larger than $G\mu=5\times10^{-8}$, which are already at the limit of being excluded. 

Foregrounds for such detection are formidable, but they are smoothly varying in frequency and if they can be fit to a low degree polynomial the analysis in ref.~\cite{Pritchard:2010pa} shows that we may be able to measure a 100 mK signal with a sigma of about 20 mK. This would allow us to see a WF trough in one part of the sky and distinguish it from a cosmic string WF trough in another part of the sky. 

\acknowledgments
I would like to thank Robert Brandenberger and Gil Holder for useful discussions. 
This work was supported by the FQRNT Programme de recherche 
pour les enseignants de coll\`ege.


\end{document}